# Physics-AI Symbiosis

Bahram Jalali*, Achuta Kadambi and Vwani Roychowdhury
Electrical and Computer Engineering Department, UCLA
*jalali@ucla.edu

## Abstract
The phenomenal success of physics in explaining nature and designing hardware is predicated on efficient computational models. A universal codebook of physical laws defines the computational rules and a physical system is an interacting ensemble governed by these rules. Led by deep neural networks, artificial intelligence (AI) has introduced an alternate end-to-end data-driven computational framework, with astonishing performance gains in image classification and speech recognition and fueling hopes for a novel approach to discovering physics itself.  These gains, however, come at the expense of interpretability and also computational efficiency; a trend that is on a collision course with the expected end of semiconductor scaling known as the Moore's Law. With focus on photonic applications, this paper argues how an emerging symbiosis of physics and artificial intelligence can overcome such formidable challenges, thereby not only extending the latter's spectacular rise but also transforming the direction of physical science.

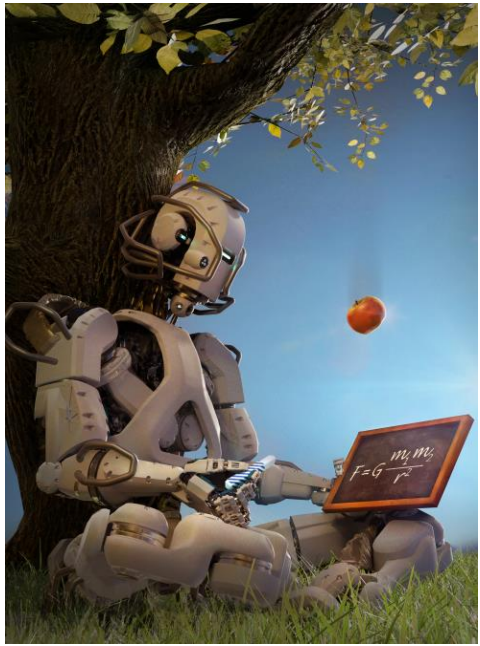

## Introduction
In the midst of the race to put a man on the moon in the 1960s, the Kalman filter performed the crucial task of estimating the trajectory of the Apollo spacecraft to the moon and back. With crude computers and imperfect sensors and with the lives of astronauts at stake, neither physics models nor sensor data alone could be relied on for accurate navigation. The Kalman filter is an early example of how blending physical and data-driven models can solve critical problems in resource constrained and data starved applications [1] [2].

## Motivation

Today, artificial intelligence (AI) and in particular deep neural networks (DNN) are revolutionizing computer vision and natural language processing (NLP). For some problems, like classifying objects in images, neural networks can even outperform humans [3] [4]. Remarkably, these deployed AI algorithms are able to succeed without a priori structures built into their designs; instead, they leverage billions of free parameters that adapt to data. Unfortunately, learning the correct setting of these free parameters means that the contemporary paradigm of neural networks requires massive datasets and long training times. These free parameters constitute a black box and it is difficult for such neural networks to offer interpretability and robustness guarantees. Furthermore, current trajectory of neural networks where performance gains are obtained by increase in network size is further challenged by the fundamental scaling limits of the semiconductor technology.

This paper explores emerging techniques in which physics is blended with artificial intelligence to mitigate these problems and create new capabilities. More specifically, we see three emerging categories: (1) Inverse Design where specialized neural networks design hardware systems; (2) Nature as a computer, where physical proxies are used to accelerate computations; and finally (3) Nature blends with computers, where the laws of physics are used as priors to enhance neural networks, and neural networks are explored to potentially learn new physics. Most of the examples described in this paper are concerned with optical physics and applications to photonic systems.

## 1. Inverse Design and Optimization

This section discusses the application of machine learning to improve the design and functionality of optical instruments. To create today's nanometer-scale integrated circuits, diffraction induced distortion in imaging of ever-shrinking circuit features must be compensated by altering the photomask design. Conventional mask optimization technique, known as Optical Proximity Correction (OPC), compensate for distortions by pre-distorting the desired circuit patterns before printing them onto the wafer. Inverse lithography is a powerful approach to OPC which minimizes the error between the printed image and the target, subject to constraints of lithography imposed by optical physics. While OPC is effective, it is computationally expensive in state-of-the-art semiconductor technology nodes. Neural network-based designs of photolithography masks offer an alternative or complementary solution [5]. For example, recent work has shown how a Generative Adversarial Network (GAN) can be utilized for mask design [6]. GANs are deep neural network architectures comprised of two competing networks [7]. One network, termed the generator, synthesizes new patterns while a second network, known as the Discriminator, attempts to distinguish synthesized target patterns from known ones. In applying this concept to mask design, the generative model was initialized with a pre-training that is guided by inverse lithography, i.e., the physical model. The generator is an auto-encoder consisting of an encoder and decoder subnets. The encoder extracts a hierarchical layout feature while, operating in reverse, the decoder predicts the pixel level mask corrections. To ensure one-on-one correspondence, the Generator should be able to deceive the Discriminator only if the generated mask precisely produces the desired (target) layout on the wafer. To achieve this, the classifier predicts positive or negative labels when presented a target-mask pair.

Any process employed in semiconductor manufacturing must be extremely high yield to meet the stringent economics of this industry. Given the high fabrication cost of advanced technology nodes, the black-box nature of neural networks and challenges associated with the lack of interpretability may need to be addressed before this technology is utilized in manufacturing. While additional data is needed to show neural network-based techniques can achieve high manufacturing yields and robustness, the tremendous potential economic payoff will nevertheless fuel research and development activities in this fascinating area.

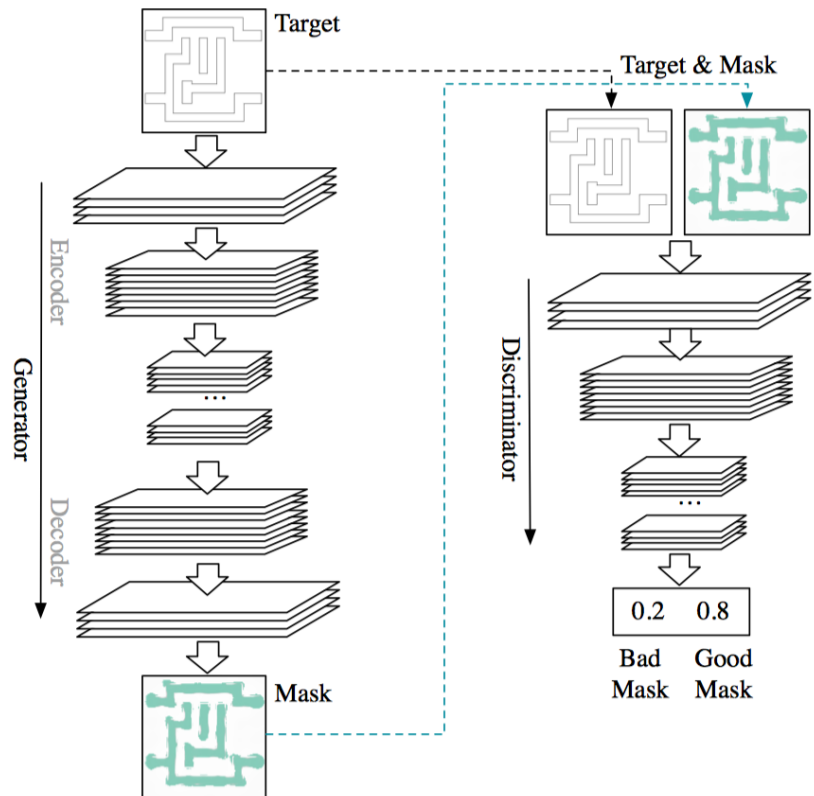

Figure 1. Neural network based optical proximity correction for photolithography mask design [6]. The objective is to correct for distortions caused by optical diffraction and reproduce on the wafer the original layout drawn by the circuit designer. A generative adversarial network (GAN) consisting of Generator and Discriminator subnets learns the mapping between the target circuit pattern and the mask. The trained model then takes target pattern as input and generates the corresponding mask.

Machine learning is also a driving force in the design of photonic instrumentation. For example, it is possible to use machine learning to peer into the complex optical dynamics that occur when femtosecond pulses propagate through nonlinear optical media such as a photonic crystal fiber. Such systems exhibit fascinating dynamics, the best-known example of which is the optical rogue wave; very rare but extreme flashes of broadband light that were discovered using time stretch data acquisition [8] [9]. The application of machine learning has made it possible to obtain the statistics of the temporal pulse trains associated with optical rogue wave events, from the spectral magnitude [10].

## 2. Nature as a Computer

So far, we have discussed the use of AI to design photonic systems. Unfortunately, in many cases the existing AI algorithms are often computationally intensive. The second theme of this paper pertains to the use of analog systems to efficiently perform specialized computations, or transform the data in a manner that reduces the burden on the digital processor. Here, a natural system either serves as an analog hardware accelerator for the digital computer [11]. One example is an ultrafast nonlinear optical waveguide where dynamics occur much more rapidly than they can be simulated by a digital computer. In this case, the properties of an inaccessible and complex system, such as hydrodynamic phenomena, are simulated with a compact proxy system. A natural computer based on such rapid dynamics can be a proxy for the computation of fluid dynamics phenomena. With an instrument capable of capturing the output in realtime, billions of outcomes resulting from complex nonlinear interactions can be readily acquired in milliseconds whereas numerical simulations could take days or longer. In other words, the natural proxy system computes many orders of magnitude more rapidly than a digital computer.

Despite the increasing speed of digital processors, the execution time of AI algorithms are still orders of magnitude slower than the time scales in ultrafast optical imaging, sensing, and metrology. To address this problem, a new concept in hardware acceleration of AI that exploits femtosecond pulses for both data acquisition and computing has recently been demonstrated [12] [13]. In the experiments, data is first modulated onto the spectrum of a supercontinuum laser. Then, a nonlinear optical element performs a data transformation analogous to a kernel operation projecting the data into an intermediate space in which data classification accuracy is enhanced. The output spectrum is sampled by a spectrometer and is sent into a digital classifier that is lightly trained. We show that the nonlinear optical kernel can improve the linear classification results similar to a traditional numerical kernel (such as the radial-basis-function) but with orders of magnitude lower latency. We further show that this technique can work with various other digital classifiers. Finally, we demonstrate that the technique is resilient to nonidealities such as additive and quantization noise in the system. Presently, the performance is data-dependent due to the limited degrees of freedom and the unsupervised nature of the optical kernel.

Wave propagation in a metamaterial can be exploited to perform specialized computational tasks such as solving integral equations [14]. Integral equations are ubiquitous in science and engineering and are described by a kernel function which performs a specific transformation of the input function. To create the metamaterial for solving a specific integral equation inverse-problem optimization approach was used to identify the structure that exhibits the desired computational kernel.

A third example of using nature as the substrate for computation is the use of a network of mechanical springs to perform tasks in the field of computer vision. The aim is to construct a geometric model of specific categories of objects: each node is a view of a particular part, shared by a majority of the exemplars of the given category, and each edge represents springs with different stiffness parameters to capture pairwise geometric constraints. Recent methods [15] have shown how such spring models can be learned in an automated manner from much fewer learning examples than conventional supervised learning methods. This geometric and explainable model (instead of a black-box model) mimics the way the human brain is theorized to represent object categories that are learned from only positive examples, and uses a natural model (i.e., elastics) as a basis for the computing architecture.

A fourth example pertains to a large body of interesting research centered on optical emulation of neural network computation. The goal is to use lasers and masks to emulate the architecture of a neural network. This is believed to enable machine learning to occur at the "speed of light". Specific approaches include the use of spatial light modulators for matrix manipulation or using integrated optical circuits. These interesting techniques attempt to mimic the functions of a neural network in analog optics, using masks, LCD panels or electrooptic modulation [16] [17].

In addition to physical systems acting as natural computers, principles of physics can be used to design computationally efficient algorithms for computing and solving problems in multiple domains. In computer vision, for example, a highly successful algorithm for image denoising is based on anisotropic diffusion, a non-linear and space-variant generalization of the regular diffusion process. The algorithm performs content-aware smoothing, leading to noise reduction while simultaneously preserving fine details [18]. The phase stretch transform (PST) is a computationally efficient edge and texture detection algorithm for visually impaired images [19] [20]. Inspired by the physics of the photonics time-stretch technique, the algorithm transforms the image by emulating near-field propagation of light through a medium with engineered dispersive/diffractive property followed by coherent detection. The algorithm has been effective in improving the resolution of MRI images [21], in single molecule imaging and in identifying and tracking of live cells, and automated creation of cell lineage trees from time lapse microscopy.

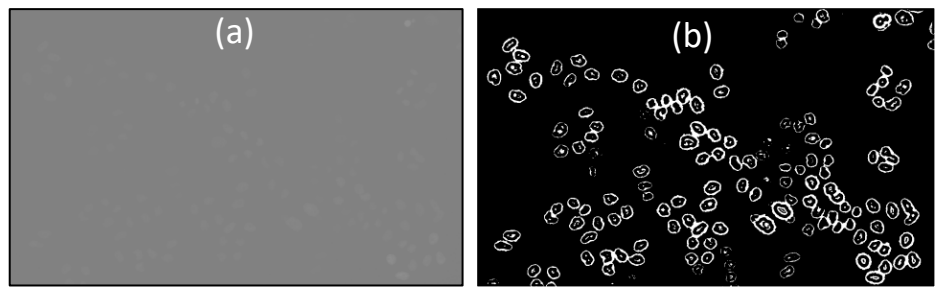

*Figure 2. A physics-inspired live-cell detection algorithm. (a) Image of live cells under a microscope [http://celltrackingchallenge.net/]. The content of cells is mostly water hence they are nearly transparent and hardly visible unless they are stained with chemicals. (b) The same image after application of contrast enhancement plus Phase Stretch Transform (PST). Cell boundaries become clearly visible. The algorithm emulates near field diffraction followed by coherent (phase) detection* [19]-[20]. *The algorithm has been integrated with an adaptive and computationally-light learning framework leading to best-in-class resolution enhancement in visually impaired brain MRI images* [21].

## 3. Nature Blends with Computer

Filling the gap between purely physical models and empirical data-driven approaches, the physics-based learning is currently undertaking a quest for a unified approach that can blend physics and learning. Such an approach would be able to handle variations in the quality of the physical prior and the dataset, enabling the algorithm to generalize across a range of physical problems. Recent work has made steps in that direction by identifying the optimal architecture by searching over a topological space of neural network configurations. This algorithm, dubbed

PhysicsNAS (Physics Neural Architectural Search), is able to blend physics and learning for problems such as collision estimation and projectile motion. While, current work only considers elementary kinematic tasks, the future of physics-based learning lies in its applicability to handle difficult and partially defined physical models, and potentially the ability to discover the underlying equations behind them [22]. Using neural networks and physical insights, computationally difficult tasks, such as the three-body problem, can be solved [23].

A second theme in blending nature and the computer is in the design of optical "computational sensing" systems. The increasingly popular field of "computational imaging" seeks to jointly design the camera system with post-processing algorithms. By blending the design of the natural and computing system, we begin to observe unconventional, new designs emerging. For example, one can rethinks the design of the color filter (ordinarily inspired by a crude model of human eyesight) to the new context of optimizing the color filter array for high-level AI tasks in computer vision [24]. This computational imaging methodology has also been studied in the realm of microscopy where the coded illumination patterns are designed with physical constraints in mind [25]. Here, an LED microscope uses illumination patterns that are computationally chosen given optical phase constraints. These themes are closely related to the use of AI models to design camera hardware, but involve hybrid physics-based AI algorithms as well.

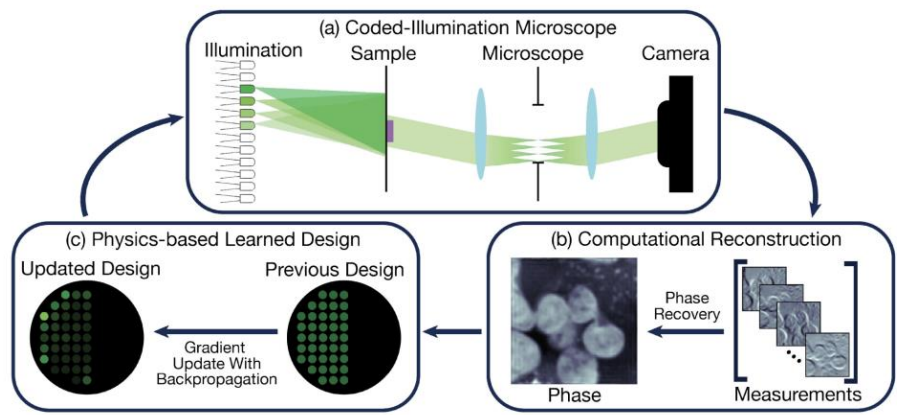

*Figure 3. An LED-array microscope captures intensity measurements with coded illumination patterns* [25]. *The illumination patterns are updated using a gradient descent framework that takes phase recovery into account.*

Yet another emerging trend in blending nature and learning, is to use DNNs as an integral part of physics models. Recognizing that computing the transmission spectra of metasurfaces given its geometric pattern involves long computing times necessary to solve the associated Maxwell equations, it is possible to train a DNN to simulate the physical system [26]. Once trained, the DNN is used as the physics model, capable of replacing the Maxwell equations-based representation, and obviates the need for using complex numerical methods. Similarly, high-fidelity modeling of unsteady flows is one of the major challenges in computational physics, with diverse applications in engineering. Recent work has shown that instead of modeling such flow with Navier-Stokes equations, and then solving them, one can directly train a DNN with training datasets from real life experiments [27]. Once trained, the DNNs provide high fidelity predictions for new situations.

Yet another ambitious theme of blending nature with computers is an emerging area known as "discovering physics". The aim is to design artificial intelligence algorithms that can discover natural laws. For instance, if a machine were to observe a video of a physical event (e.g., an apple falling), would it be possible for the machine to obtain the symbolic expression for projectile motion. The problem has thus far been partially solved. If a machine were to have an understanding of certain natural laws, it is possible to mine videos for the rest of laws and discover physical equations [28]. However, this algorithm requires some knowledge of a physical foundation. For example, it can obtain the equation for projectile motion, but the existence and calculation of velocity must be provided as inputs.

A different approach to blending physics with AI concerns the use of two neural networks, one modeling the evolution through the physical system and the other providing a prior in the form of an approximate solution [29]. This deep learning framework learns the underlying dynamics and predicts the system's future states. It is an interesting and a bold approach that can lead to automated discoveries of PDEs that govern physical laws.

More recent work has aimed to offer a more complete picture of physical discovery. For example, a special neural network can be trained to discover physical laws from video without prior human intervention, by creating a pipeline that extends a latent embedding module based upon variational auto-encoders [30]. Figure 4 compares how humans discover physical laws versus how machines discover physical laws.

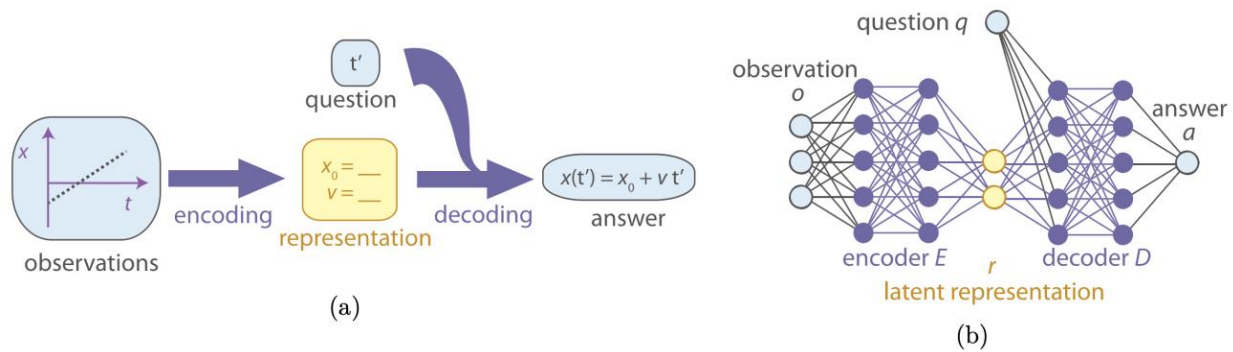

Figure 4. Imagine if it was possible to discover new laws of imaging physics. Recent work in AI has created deep learning models that can potentially discover laws of imaging physics. Shown here is the contrast between how (a) humans discover natural laws with (b) how machines may one day discover such laws [31]. Humans discover physical laws by representing a set of observations with a compact representation. In contrast, machines compress massive, high-dimensional amounts of data into a much more compact latent representation.

## Conclusion

The power of deep neural networks stems from having billions of free parameters that adapt to data. The training of these parameters requires massive datasets and being a black box, these models lack interpretability and robustness guarantees. Applications in realtime control such as autonomous vehicles and drones have additional strict requirements for low latency. Concurrently, the impressive performance gain in neural networks has come at the expense of exponential growth in network size with the concomitant growth in computational complexity, memory and power consumption [32]. Yet, the semiconductor technology is reaching fundamental scaling limits, a trend generally referred to as "the end of Moore's law". These trends call for a fresh look into the design of artificial intelligence systems.

The emerging symbiosis of physics including optical physics and data science in general and neural networks in particular, is evolving along three directions. Neural networks are being utilized for inverse design of optimized physical systems such as lithography masks. In Nature as a Computer, physical phenomena are being used to accelerate digital computing through analog feature extraction and data representation. Physics is blending intricately with computation for applications such as the joint design of a camera system with post-processing algorithms. Such an integration allows neural networks to be both data and power efficient, while allowing interpretability, and addressing some of the scaling and power consumption challenges. Conversely, such an integration has the potential to provide reciprocal gains in physics, and an ambitious but a potentially high impact recent undertaking is the attempt to discover new physical laws from data without human intervention. While such visionary projects are in their infancy, the symbiosis of these separate fields will surely transform the vector of physics and optics in new and unexpected ways.

# Bibliography

[1]  R. Kalman, "A new approach to linear filtering and prediction problems," *Journal of Basic Engineering,* vol. 82, no. 1, pp. 35-45, 1960.

[2]  T. Kailath and P. Frost, "An innovations approach to least-squares estimation--Part II: Linear smoothing in additive white noise," *IEEE Transactions on Automatic Control,* vol. 13, no. 6, pp. 655-660, 1968.

[3]  A. Krizhevsky, I. Sutskever and G. Hinton, "Imagenet classification with deep convolutional neural networks," in *Conference on Neural Information Processing Systems*, 2012.

[4]  K. He, X. Zhang, S. Ren and J. Sun, "Delving deep into rectifiers: Surpassing human-level performance on imagenet classification," in *Proceedings of the IEEE International Conference on Computer Vision*, 2015.

[5]  X. Ma, Q. Zhao, H. Zhang, Z. Wang and G. Arce, "Model-driven convolution neural network for inverse lithography," *Optics Express,* vol. 26, no. 25, pp. 32565-32584, 2018.

[6]  H. Yang, S. Li, Y. Ma, B. Yu and E. Young, "GAN-OPC: Mask optimization with lithography-guided generative adversarial nets," in *ACM/ESDA/IEEE Design Automation Conference (DAC)*, 2018.

[7]  I. Goodfellow, J. Pouget-Abadie, M. Mirza, B. Xu, D. Warde-Farley, S. Ozair, A. Courville and Y. Bengio, "Generative adversarial nets," in *Advances in Neural Information Processing Systems*, 2014.

[8]  D. R. Solli, C. Ropers, P. Koonath and B. Jalali, "Optical rogue waves," *Nature,* vol. 450, no. 7172, pp. 1054-1057, 2007.

[9]  A. Mahjoubfar, D. V. Churkin, S. Barland, N. Broderick, S. K. Turitsyn and B. Jalali, "Time stretch and its applications," *Nature Photonics,* vol. 11, no. 6, pp. 341-351, 2017.

[10] M. Narhi, L. Salmela, J. Toivonen, C. Billet, J. M. Dudley and G. Genty, "Machine learning analysis of extreme events in optical fibre modulation instability," *Nature Communications,* vol. 9, no. 1, p. 4923, 2018.

[11] D. R. Solli and B. Jalali, "Analog optical computing," *Nature Photonics,* vol. 9, no. 11, pp. 704-706, 2015.

[12] T. Zhou, F. Scalzo and B. Jalali, *Nonlinear Schrodinger Kernel Computing for Single-shot Data Acquisition and Inference,* arXiv preprint arxiv: 2012.08615.2020, 2020.

[13] B. Jalali, T. Zhou and F. Scalzo, "Time Stretch Computing for Ultrafast Single-shot Acquisition and Inference," in *Optical Fiber Communication Conference*, 2021.

[14] N. Estakhri, B. E. Mohammadi and N. Engheta, "Inverse-designed metastructures that solve equations," *Science,* vol. 363, no. 6433, pp. 1333-1338, 2019.

[15] L. Chen, S. Singh, T. Kailath and V. Roychowdhury, "Brain-inspired automated visual object discovery and detection," *Proceedings of the National Academy of Sciences,* vol. 116, no. 1, pp. 96-105, 2019.

[16] X. Lin, Y. Rivenson, N. T. Yardimci, M. Veli, Y. Luo, M. Jarrahi and A. Ozcan, "All-optical machine learning using diffractive deep neural networks," *Science,* vol. 361, no. 6406, pp. 1004-1008, 2018.

[17] J. K. George, A. Mehrabian, R. Amin, J. Meng, T. Ferreira De Lima, A. N. Tait, B. J. Shastri, T. El-Ghazawi, P. R. Prucnal and V. J. Sorger, "Neuromorphic photonics with electro-absorption modulators.," *Optics Express,* vol. 27, no. 4, pp. 5181-5191, 2019.

[18] P. Perona and J. Malik, "Scale-space and edge detection using anisotropic diffusion," *IEEE Transactions on pattern analysis and machine intelligence,* vol. 12, no. 7, pp. 629-639, 1990.

[19] M. H. Asghari and B. Jalali, "Edge detection in digital images using dispersive phase stretch transform," *International journal of biomedical imaging,* 2015.

[20] M. Suthar, H. Asghari and B. Jalali, "Feature enhancement in visually impaired images," *IEEE Access,* vol. 6, pp. 1407-1415, 2017.

[21] S. He and B. Jalali, "Fast Super-Resolution in MRI Images Using Phase Stretch Transform, Anchored Point Regression and Zero-Data Learning," in *IEEE International Conference on Image Processing*, 2019.

[22] M. Schmidt and H. Lipson, "Distilling free-form natural laws from experimental data," *Science,* vol. 324, no. 5923, pp. 81-85, 2009.

[23] P. G. Breen, C. N. Foley, T. Boekholt and S. P. Zwart, "Newton vs the machine: solving the chaotic three-body problem using deep neural networks," in *arXiv preprint arXiv:1910.07291*, 2019.

[24] A. Chakrabarti, "Learning sensor multiplexing design through back-propagation," in *Advances in Neural Information Processing Systems*, 2016.

[25] M. Kellman, E. Bostan, N. Repina and L. Waller, "Physics-based learned design: Optimized coded-illumination for quantitative phase imaging," *IEEE Transactions on Computational Imaging,* 2019.

[26] Z. Liu, D. Zhu, S. P. Rodrigues, K.-T. Lee and W. Cai, "Generative model for the inverse design of metasurfaces," *Nano letters,* vol. 18, no. 10, pp. 6570-6576, 2018.

[27] J. N. Kutz, "Deep learning in fluid dynamics," *Journal of Fluid Mechanics,* vol. 814, pp. 1-4, 2017.

[28] S. Huang, Z.-Q. Cheng, X. Li, X. Wu, Z. Zhang and A. Hauptmann, "Perceiving Physical Equation by Observing Visual Scenarios," in *arXiv preprint arXiv:1811.12238*, 2018.

[29] M. Raissi, "Deep hidden physics models: Deep learning of nonlinear partial differential equations," *Journal of Machine Learning Research,* vol. 19, no. 1, pp. 932-955, 2018.

[30] P. Chari, C. Talegaonkar, Y. Ba and A. Kadambi, "Visual Physics: Discovering Physical Laws from Videos," in *arXiv preprint arXiv:1911.11893*, 2019.

[31] R. Iten, T. Metger, H. Wilming, L. D. Rio and R. Renner, "Discovering physical concepts with neural networks," in *arXiv preprint arXiv:1807.10300*, 2018.

[32] X. Xu, Y. Ding, S. Hu, M. Niemier, J. Cong, Y. Hu and Y. Shi, "Scaling for edge inference of deep neural networks.," *Nature Electronics,* vol. 1, no. 4, pp. 216-222, 2018.